\documentclass[conference]{IEEEtran}
\IEEEoverridecommandlockouts
\usepackage{cite}
\usepackage{amsmath,amssymb,amsfonts}
\usepackage{algorithmic}
\usepackage{graphicx}
\usepackage{caption}
\usepackage{multirow}
\usepackage{booktabs}
\usepackage{textcomp}
\usepackage{xcolor}
\usepackage{diagbox}
\usepackage{multirow}
\usepackage{multirow}
\usepackage{graphicx}
\usepackage{colortbl}
\usepackage{pifont}
\usepackage{hhline}
\usepackage{threeparttable}
\usepackage{float}  %设置图片浮动位置的宏包
\usepackage{subfigure}  %插入多图时用子图显示的宏包
\usepackage{graphicx}  %插入图片的宏包
\def\BibTeX{{\rm B\kern-.05em{\sc i\kern-.025em b}\kern-.08em
    T\kern-.1667em\lower.7ex\hbox{E}\kern-.125emX}}
\begin{document}

\title{A Multi-Task Learning Model for Super Resolution of Wireless Channel Characteristics
}
\author{\IEEEauthorblockN{Xiping Wang\IEEEauthorrefmark{1}, Zhao Zhang\IEEEauthorrefmark{1}, Danping He\IEEEauthorrefmark{1}, Ke Guan\IEEEauthorrefmark{1}\IEEEauthorrefmark{2}, Dongliang Liu\IEEEauthorrefmark{3}, Jianwu Dou\IEEEauthorrefmark{3},
and Bo Sun\IEEEauthorrefmark{3}}

\IEEEauthorblockA{\IEEEauthorrefmark{1}State Key Laboratory of Rail Traffic Control and Safety, Beijing Jiaotong University, 100044, Beijing, China}

\IEEEauthorblockA{\IEEEauthorrefmark{2}Frontiers Science Center for Smart High-speed Railway System, 100044, Beijing, China}

\IEEEauthorblockA{\IEEEauthorrefmark{3}State Key Laboratory of Mobile Network and Mobile Multimedia Technology, 518055, Shenzhen, China}

Corresponding Author: Danping He (e-mail: hedanping@bjtu.edu.cn).

% \author{\IEEEauthorblockN{Xiping Wang$^{1}$, Zhao Zhang$^{1}$, Danping He$^{12*}$, Ke Guan$^{12}$, Zhangdui Zhong$^{12}$, Dongliang Liu$^{34}$}
% \IEEEauthorblockA{\textit{State Key Laboratory of Railway Traffic Control and Safety, Beijing Jiaotong University, Beijing, China} \\

% \textit{State Key Laboratory of Railway Traffic Control and Safety, Beijing Jiaotong University, Beijing, China$^{1}$}\\
% \textit{Frontiers Science Center for Smart High-speed Railway System, Beijing, China$^{2}$}\\
% \textit{State Key Laboratory of Mobile Network and Mobile Multimedia Technology, Shenzhen, Guangdong, China$^{3}$}\\
% \textit{ZTE Corporation, Shenzhen, Guangdong, China$^{4}$}\\
% \text{\{16211133,18211026,danpinghe,kguan,zhdzhong\}@bjtu.edu.cn$^{12}$, }
% \text{liu.dongliang1@zte.com.cn$^{34}$}
% }

}

\maketitle

\begin{abstract}
Channel modeling has always been the core part in communication system design and development, especially in 5G and 6G era. Traditional approaches like stochastic channel modeling and ray-tracing (RT) based channel modeling depend heavily on measurement data or simulation, which are usually expensive and time consuming. In this paper, we propose a novel super resolution (SR) model for generating channel characteristics data. The model is based on multi-task learning (MTL) convolutional neural networks (CNN) with residual connection. Experiments demonstrate that the proposed SR model could achieve excellent performances in mean absolute error and standard deviation of error. Advantages of the proposed model are demonstrated in comparisons with other state-of-the-art deep learning models. Ablation study also proved the necessity of multi-task learning and techniques in model design. The contribution in this paper could be helpful in channel modeling, network optimization, positioning and other wireless channel characteristics related work by largely reducing workload of simulation or measurement. 

\end{abstract}

\begin{IEEEkeywords}
Wireless channel modeling, ray-tracing (RT), super resolution (SR), multi-task learning (MTL), convolutional neural network (CNN)
\end{IEEEkeywords}

\section{Introduction}
Thanks to the fast advances in the fifth generation (5G) wireless communication, our world is stepping into the era of Internet of Everything (IoE)\cite{RN62}. To enable ultra-low latency and high reliability communication services in dense connected areas like urban districts, accurate wireless channel model is a necessity. Correct knowledge of the propagation channel is also critical for radio coverage estimation, network optimization and for many related applications\cite{RN58}. 
Naturally, channel modeling is seen as the foundation for planning and optimizing communication and related systems\cite{RN8}.  

Channel modeling is the process of characterizing the propagation principles of radio waves in realistic environments, and provides insight theoretical guidance for the design, deployment and optimization of communication systems. Generally, stochastic channel modeling (SCM) and ray-tracing (RT) based deterministic modeling are two main modeling approaches\cite{RN57}. For SCM, channel characteristics like path loss (PL), propagation condition (line of sight or non line of sight), delay spread, angular spreads and Rician K-factor are required to generate channel coefficients and thereafter to model the wireless channel. Massive channel measurements in different environments, which are usually time-consuming and expensive, must be conducted to obtain the necessary channel characteristics data\cite{RN63}. On the other side, RT based modeling approach can generate accurate channel data if given precise propagation environment and configuration, but at the cost of high computational complexity and enormous calculation time\cite{9135220}. Therefore, a fast and reliable channel characteristics data generation method will effectively address these limitations.

\begin{figure*}[t]
\centering
\noindent
  \includegraphics[width=6.5in]{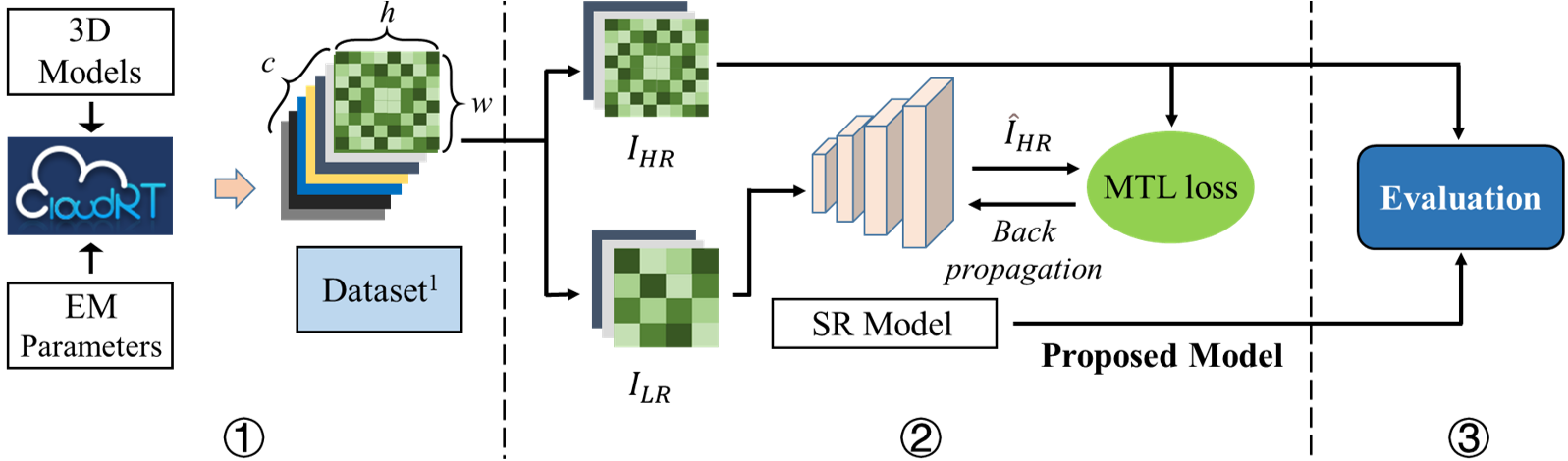}
  \caption{The overview of proposed MTL SR model for wireless channel characteristics. \ding{172} RT simulation and dataset generation. \ding{173}  The SR model training process.  \ding{174} Evaluation.  \\ Dataset$^{1}$: Channel characteristics dataset}\label{overview}
\centering
\end{figure*}

One of the promising solutions is machine learning (ML) method. The recent surge of ML is revolutionizing almost every branch of science and technology, including wireless channel modeling. The wireless channel is a time-varying nonlinear system, which contains multi-dimensional information in time, spatial and frequency domain. Machine learning has very powerful learning and inferring capabilities. It can automatically learn from channel data so that the structural relationship between data in complex environments can be extracted to approximate nonlinear systems. Moreover, ML is very efficient in mining information from high dimensional data, which can significantly expedite data processing. 

Current research in this area are still inadequate. Most of recent works focused on predicting only one of the channel characteristics like PL (Radio Coverage Prediction)\cite{RN58}. The input is mostly restricted to building information, satellite image\cite{RN65}, map\cite{RN26} and just one of the characteristics. Few works attempt to estimate several channel characteristics at the same time. As for ML methods, traditional algorithms such as random forests (RF), support vector machine (SVM) and K-nearest neighbors (KNN) as well as deep learning (DL) models such as convolutional neural network (CNN)\cite{RN26}, Transformer \cite{RN25} and generative adversarial network (GAN) \cite{RN55} are frequently employed. Also, the estimation targets are usually restricted to one or two characteristics. In general, few works have studied multi-task learning (MTL) models for generating channel characteristics data with several characteristics as input. The connection between different characteristics awaits to be exploited.

In this paper, instead of prediction or estimation, we propose a super resolution (SR) model for channel characteristics. The model is based on MTL CNN with residual connection. Overview of our work is shown in Fig.\ref{overview}. Given 3D models of urban areas and the corresponding EM parameters, CloudRT platform outputs channel characteristics dataset and the dataset is used for SR model training. Data from the  dataset are degraded into low resolution data as input. Original high resolution data are used as the ground truth. MTL loss is employed to better balance multi tasks. We evaluate our proposed SR model by ablation study and comparisons with other DL models. Specifically, we make the following contributions:
\begin{itemize}

\item In dense urban areas, RT simulation by self developed CloudRT was conducted and channel characteristics dataset are constructed based on simulation results.

\item A residual network based MTL SR model is proposed. Weighted masks are added in loss function. Homoscedastic uncertainty is employed to balance the single task losses during training. Residual connection and \textit{iterative up-and-down} technique are implemented in CNN blocks for better SR performances.

\item Training process and results of evaluation are provided. The proposed SR approach generally performs better than other state-of-the-art DL models. Comparing with baseline, the proposed approach could achieve very good SR results in all channel characteristics targets and deteriorate significantly less than baseline with larger scale factor. Ablation study proved that the techniques in training and model design are necessary.

\end{itemize}

\section{Simulation and Dataset Construction}

\begin{table}[ht]
\begin{center}
\caption{RT Simulation configuration} \label{TableSimConf}
\begin{tabular}{|l|l|l|}
 \hline
 Carrier frequency & 3.55 GHz\\
 \hline
System bandwidth & 100 MHz\\
 \hline
 Frequency resolution & 1 MHz\\
 \hline
Antenna & Omni-directional vertical polarization\\
 \hline
 Simulation range & 1 km${\times}$1 km\\
 \hline
  Tx location & 30 - 50 m above the ground\\
 \hline
   Rx location & 2 m above the ground\\
 \hline
\end{tabular}
\end{center}
\end{table}

This section describes the procedures of RT simulation and dataset construction. Self developed CloudRT\cite{DanpingTutorial}\cite{OpenScience}  platform is used to generate channel characteristics data.

RT approach is widely used to generate accurate channel characteristics data in a specific environment. Based on 3D electronic map and electromagnetic (EM) parameters provided by ITU-R P.1238-7, RT simulation is conducted in dense urban areas on CloudRT platform. More than 100 simulation regions are manually selected from urban areas of four major cities in China: Beijing, Shanghai, Hangzhou and Xi'an. The simulation regions are 1km $\times$ 1km squares with 3D electronic map resolution of 200 $\times$ 200. Transmitter (Tx) is located on one of the high buildings near the center of simulation region, and a large LOS area should be maintained. The receivers (Rx) are located 2 meters above ground, uniformly distributed by distance of 5 meters on the horizontal plane. Only receivers outside buildings are considered in simulation. Table \ref{TableSimConf} summarizes the simulation configuration and details of it can be found in\cite{APS-New}.

\begin{table}[t]
\renewcommand\arraystretch{1.2}

\caption{Channel characteristics dataset} \label{TableChannelData}
\begin{center}

\begin{tabular}{l|lll}
\hline
\multicolumn{2}{l|}{\textbf{Characteristic}} & \textbf{Normal Range} & \multicolumn{1}{l}{\textbf{NaN Value}} \\ \hline
\multicolumn{2}{l|}{PL [dB]}               & [-200,0)              & 200                                                                    \\
\multicolumn{2}{l|}{$\textit{R}_p$ [dB]}  & (-30,0]               & 100                                                                   \\
\multicolumn{2}{l|}{DS [ns]}            & (0,500)               & -100                                                                   \\
\multicolumn{2}{l|}{$\phi$ [°]}           & [0,360)               & -360                                                                  \\
\multicolumn{2}{l|}{$\theta$ [°]}         & [0,180)               & -180                                                                    \\
\multicolumn{2}{l|}{LOS/NLOS}                & -1/0                  & 1                                                                        \\ \hline
\end{tabular}
\end{center}
\end{table}

\begin{equation}
\textit{R}_p=\frac{\sum_{i=1}^{N_r} P_i-P_0}{\sum_{i=1}^{N_r} P_i} 
\label{mpr}
\end{equation}

The channel characteristics dataset is constructed based on RT simulation results. Definitions of PL, root mean squared (RMS) delay spread (DS), RMS azimuth ($\phi$) angular spread of arrival, RMS elevation ($\theta$) angular spread of arrival, line of sight (LOS) and non line of sight (NLOS) follows the usual. We redefined Racian-K factor as multi-path power ratio $\textit{R}_p$ in \eqref{mpr}, where $P_i$ and $P_0$ is the power of ray $i$ and ray of direct propagation. In LOS area, $\textit{R}_p$ is the ratio of power of all rays except $P_0$ to the total power. In NLOS area, $\textit{R}_p$ is equal to zero. $\textit{R}_p$ is continuous in both LOS and NLOS areas thus suitable for ML. The 6 characteristics are also SR targets in this paper. We use PL, $\textit{R}_p$, DS, $\phi$, $\theta$, LOS/NLOS as the abbreviations or symbols of the channel characteristics and SR targets, as shown in TABLE \ref{TableChannelData}. Values which are far beyond ordinary thresholds in communication systems are set as the minimum (PL, $\textit{R}_p$) or the maximum (DS) of corresponding normal range. NaN value represents the channel characteristics data of receivers which locate inside buildings. NaN value should be void but set as real number out of normal range so that ML model is able to distinguish. A data in channel characteristics dataset is a 200 by 200 tensor with 7 channel (including building heights). In total, 753 data were generated and combined to construct the dataset. The input data are processed by down-sampling by certain scale factor and up-sampling by interpolation so that the shape remains the same.

\section{Methodology}

\subsection{Problem Definitions and Terminologies}
Super resolution is a notion of recovering high resolution (HR) data from the low resolution (LR) data\cite{SR}. The LR data mostly originates from HR data with a process of degradation:
\begin{equation}
I_{LR}=\mathcal{D}\left ( I_{HR}; \delta \right) 
\label{ILR}
\end{equation}
where $I_{LR}$, $I_{HR}$, $\mathcal{D}$, $\delta$ denote LR data, HR data, degradation process which is usually unknown, and parameters of degradation such as scaling factor. Usually $I_{LR}$, $I_{HR}$ are the input training data and the ground truth correspondingly. SR can be described as the model of recovering $I_{LR}$ to $I_{HR}$. The recovery (approximation) result is denoted as $\hat{I}_{HR}$. $\mathcal{F}$ represents the recovery model with parameters $\theta$. Take DL as the example, $\mathcal{F}$ is the neural network while $\theta$ is the weights and other related parameters in the neural network. 

\begin{equation}
\hat{I}_{HR}=\mathcal{F}\left ( I_{LR}; \theta \right) 
\label{2HR}
\end{equation}
To this end, the objective of SR is as follows:
\begin{equation}
\hat{\theta}=\mathop{\arg\min}\limits_{\theta}\mathcal{L}(\hat{I}_{HR},I_{HR})+\lambda\Phi(\theta)
\label{ObjF}
\end{equation}
where $\mathcal{L}$ is the loss function between recovery result $\hat{I}_{HR}$ and ground truth $I_{HR}$. $\Phi$ is the regularization term with parameters $\theta$ and $\lambda$ is the weight factor. The best SR model would minimize the loss to the least with respect to $\theta$.

\subsection{Residual Network based Multi-Task Learning Model}
\begin{figure}[t]
\begin{center}
\noindent
  \includegraphics[width=3.4in]{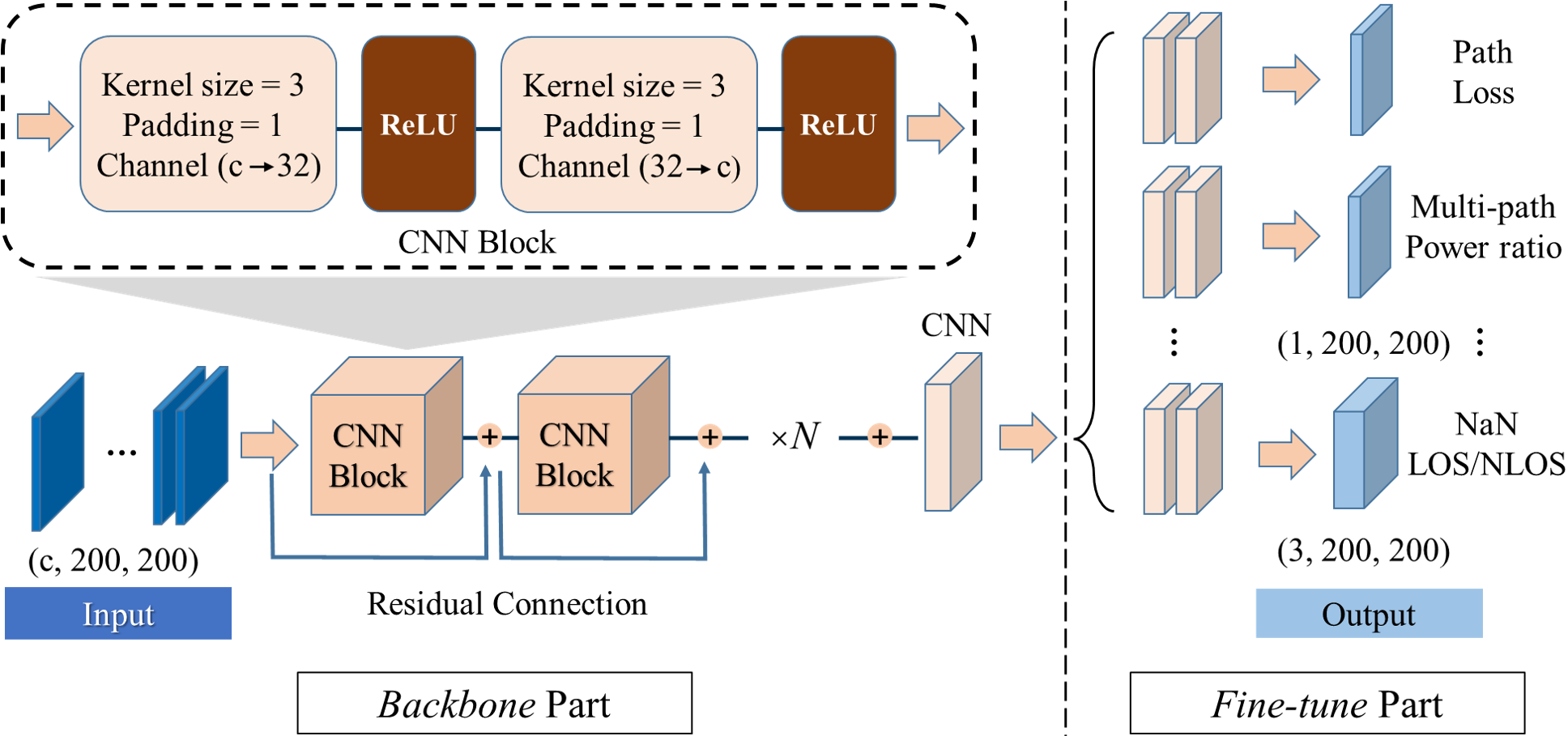}
  \caption{The overview of proposed MTL CNN model with residual connection}\label{CNNimg}
\end{center}
\end{figure}

The proposed residual network based MTL model consists of two parts: \textit{backbone} part and \textit{fine-tune} part, as in Fig. \ref{CNNimg}. The \textit{backbone} part is to extract high-dimensional features from input data while the fine-tune part focuses on each of the tasks to achieve the best SR performances.

The CNN block contains two convolutional layers with activation function ReLU. The size of data during convolution remains 200 $\times$ 200. The number of CNN blocks $N$ is set to be 3 in our work, but a larger number of blocks may also achieve excellent SR performances based on our previous experiments. To be noted, the numbers of channels in convolutional layers are set to be \textit{iterative up-and-down}, as shown in the illustration of CNN block. The \textit{iterative up-and-down} technique can filter irrelevant information in input data. Residual connection is also necessary for good SR performances. Several CNN blocks are concatenated with residual connection to construct the \textit{backbone} part. In \textit{fine-tune} part, 6 lightweight models are designed for the corresponding targets. Lightweight models are two layer CNN like in CNN block but number of channels remain fixed. Note that for NaN/LOS/NLOS, outputs of the \textit{fine-tune} nets are the probabilities of corresponding propagation conditions.

\begin{figure}[t]
\begin{center}
\noindent
  \includegraphics[width=2.8in]{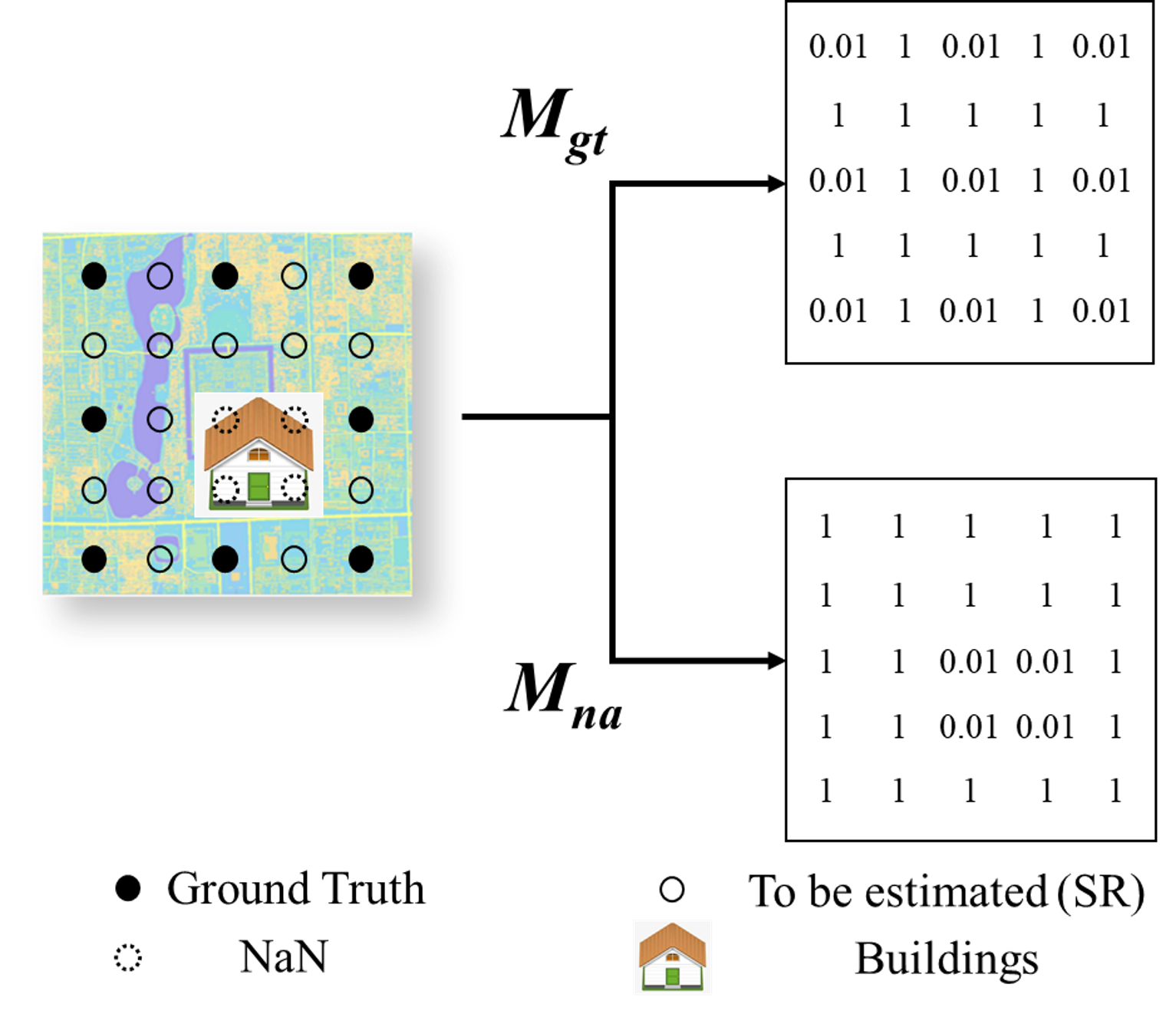}
  \caption{An illustration of $M_{gt}$ and $M_{na}$. Mask values are all set as 0.01. The background is electronic map of Tiananmen Square in Beijing, China.}\label{maskimg}
\end{center}
\end{figure}

\subsection{Loss Functions and Evaluation Metrics}

As explained in Section II, only receivers outside the building are considered in simulation. As a result, NaN values are given real number to guarantee that the input data is a regular matrix or tensor. Also, there will always be some ground truth values (elements) in the matrix of input data. However, both NaN values and ground truth values are trivial in SR process. To keep the loss function continuous and to help ML models concentrate on SR values of desired positions, weighted loss is introduced as the mask of NaN values $M_{na}$ and the mask of ground truth $M_{gt}$, which are illustrated in Fig. \ref{maskimg}. Several kinds of loss functions and evaluation metrics such as \textit{pixel loss}, and \textit{content loss} are widely used in image SR tasks. Differently, only \textit{pixel loss} is accepted in channel characteristics SR task. According to previous experiments, L1 (norm) loss is employed for training as it performs better than L2 loss and peak signal noise ratio (PSNR).

There are 6 SR targets in this paper as mentioned in section II. For LOS/NLOS, SR process is essentially to classify between LOS and NLOS area. For the rest targets, the SR process are regression. Due to the difference between classification and regression, the loss functions are categorized into two scenarios where L1 norm are used for regression and cross entropy are used for classification, as in \eqref{l1loss} and \eqref{celoss}. The loss function $\mathcal{L}_{m}$ in training can be described as follows:
\begin{equation}
\mathcal{L}_{m}(\hat{I},I)=
\begin{cases}{}
loss_{ce}(\hat{I},I),&\small{\text{if target $m$ is LOS/NLOS}}\\
loss_{l_1}(\hat{I},I),&\small{\text{others}}
\end{cases} 
\label{fullloss}
\end{equation}
where $\hat{I}$ and $I$ are SR recovered data and ground truth. Before the calculation of loss function, both $\hat{I}$ and $I$ should be weighted by \textit{Hadamard Product} with $M_{gt}$ and $M_{na}$:
\begin{equation}
I_{weighted}=I_{original}\circ M_{na} \circ M_{gt} 
\label{Hadamard}
\end{equation}

$n$, $h$, $w$, represents the number of values to be estimated (SR) and side lengths of input data. $k$ stands for the class among LOS/NLOS/NaN in \eqref{celoss}. 

\begin{equation}
loss_{l_1}(\hat{I},I)=\frac{n}{\left(hw\right) ^{2}} \sum_{i,j} |\hat{I}_{i,j}-{I}_{i,j}|_{1}
\label{l1loss}
\end{equation}

\begin{equation}
loss_{ce}(\hat{I},I)=-\frac{n}{\left(hw\right ) ^{2}} \sum_{i,j} {I}_{i,j,k}\log_{}{\hat{I}_{i,j,k}}
\label{celoss}
\end{equation}

During the pre-train stage, homoscedastic uncertainty\cite{uncertaintyweight} is employed to balance the single-task losses as in \eqref{MTLloss}. Not only the weights in neural networks $W$ but also noise parameters $\boldsymbol{\sigma}$ are trainable and updated through standard back propagation during training.

\begin{equation}
\mathcal{L}_{MTL}(W,\boldsymbol{\sigma})=\sum_{m}  \frac{\mathcal{L}_{m}}{2\sigma_{m}^{2} } +\sum_{m}log(\sigma_{m})
\label{MTLloss}
\end{equation}

The evaluation metric is much simpler than training loss. Basically, values of receiver located in building areas (NaN values) and ground truth values are not included in calculation. The metric for evaluating SR performance of LOS/NLOS is classification accuracy. For the rest targets, mean absolute error (MAE) and standard deviation of error (STDE) are regarded as the metrics.   

\section{Experiments}
\subsection{Training Configuration and Implementation}
In this work, DL training are performed by PyTorch 1.10.2 on a work station with 1 NVIDIA GeForce RTX 3090 GPU, Intel Core i9-9900K CPU and 32 GB DDR4 RAM. The training process are divided into two stage: \textit{pre-train} stage and \textit{fine-tune} stage. In \textit{pre-train} stage, \eqref{MTLloss} is used as the loss function for back propagation and weights of the entire SR model are updated. Next, in \textit{fine-tune} stage, only weights of \textit{fine-tune} part in the SR model are updated while \textit{backbone} part remains unchanged. The purpose of this design is to obtain a general feature extractor (\textit{backbone} part) for channel characteristics SR. By doing so, it's very fast and convenient to concatenate lightweight models for specific tasks when comparing with DL models aiming at only one channel characteristic. Data augmentation is employed to the training set for enhancing diversity of the inputs. The training set is transformed by rotation of 90, 180 and 270 degree as well as horizontally and vertically flipping. As a result, the training dataset is incremented by 5 times. 

The proposed SR model is trained for 100 epochs in both of two stages. The learning rate is set as 0.00001. Adam optimizer is used for gradient descent. Batchsize is set as 1. The channel characteristics dataset is randomly split into training set and test set by ratio of 7:3. Experiments with scale factor 2, 4 and 8 were conducted.

\begin{figure*}[ht]
	\centering
	\subfigbottomskip=0.5pt %设置第二行子图与第一行子图的距离，即下面的头与上面的脚的距离
	\subfigure[Path loss]{
		\label{PL_img}
		\includegraphics[width=0.32\linewidth]{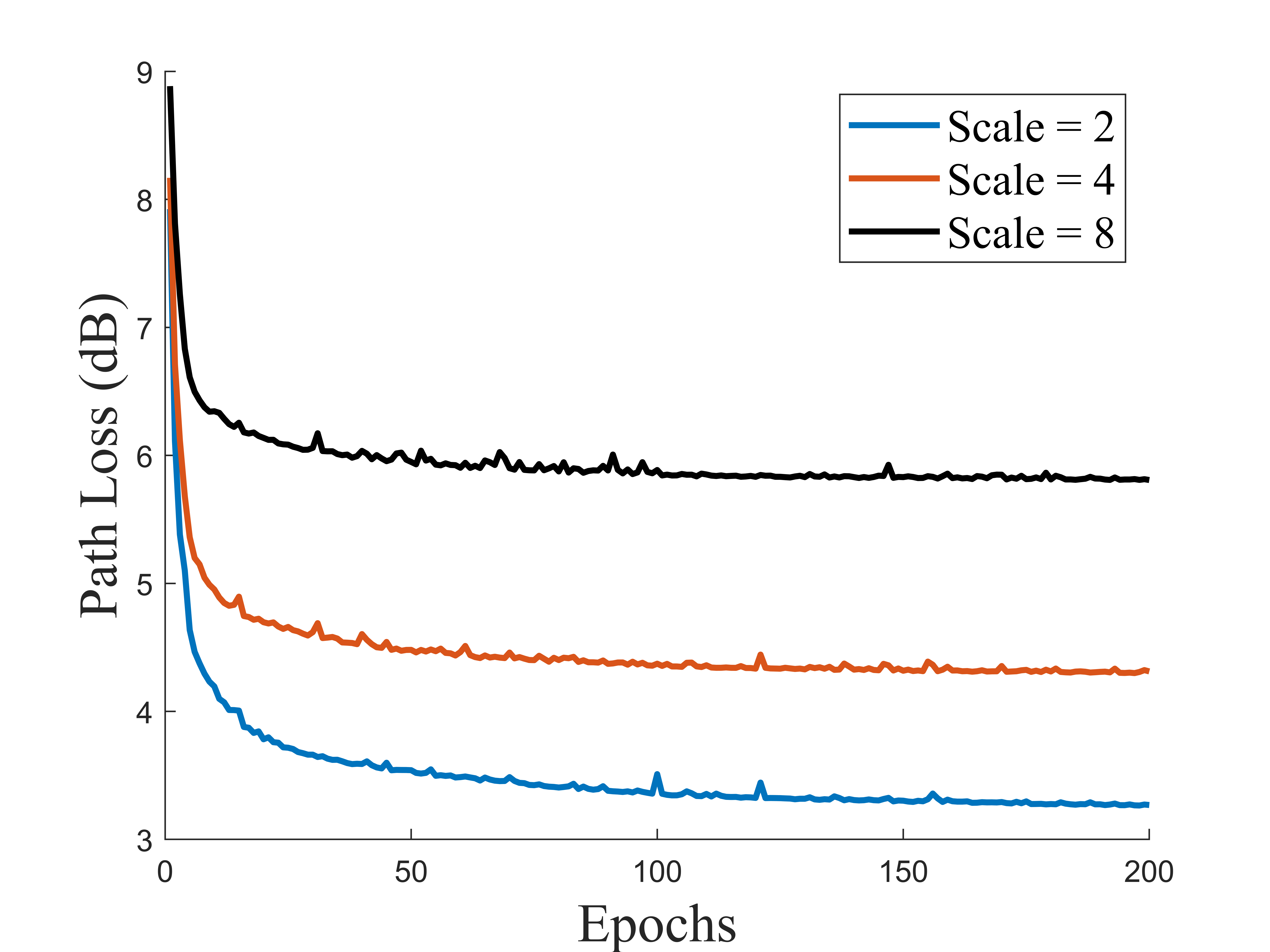}}
	\subfigure[$R_p$]{
		\label{Kp_img}
		\includegraphics[width=0.32\linewidth]{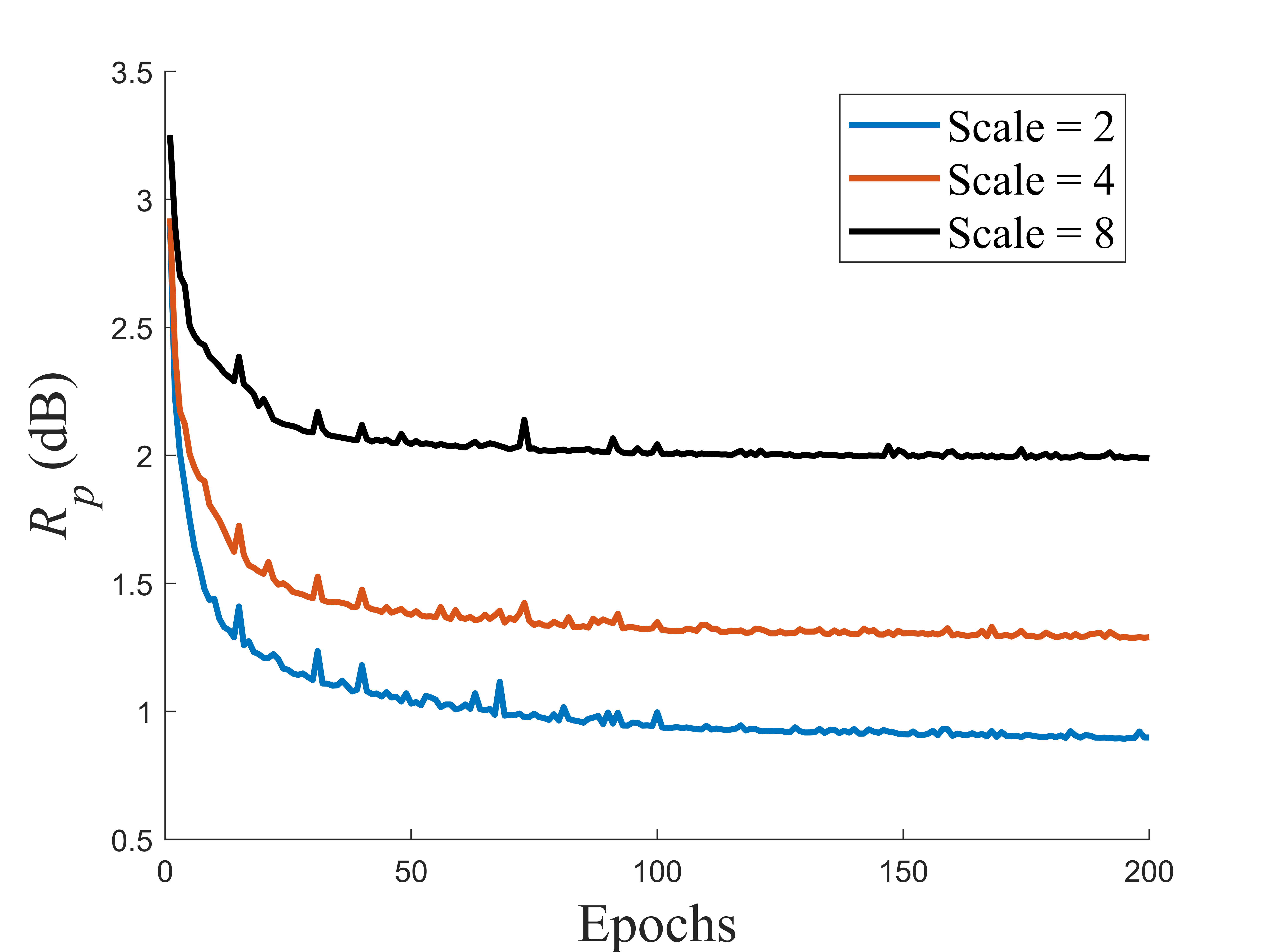}}
	\subfigure[RMS $\phi$ spread]{
		\label{phi_img}
		\includegraphics[width=0.32\linewidth]{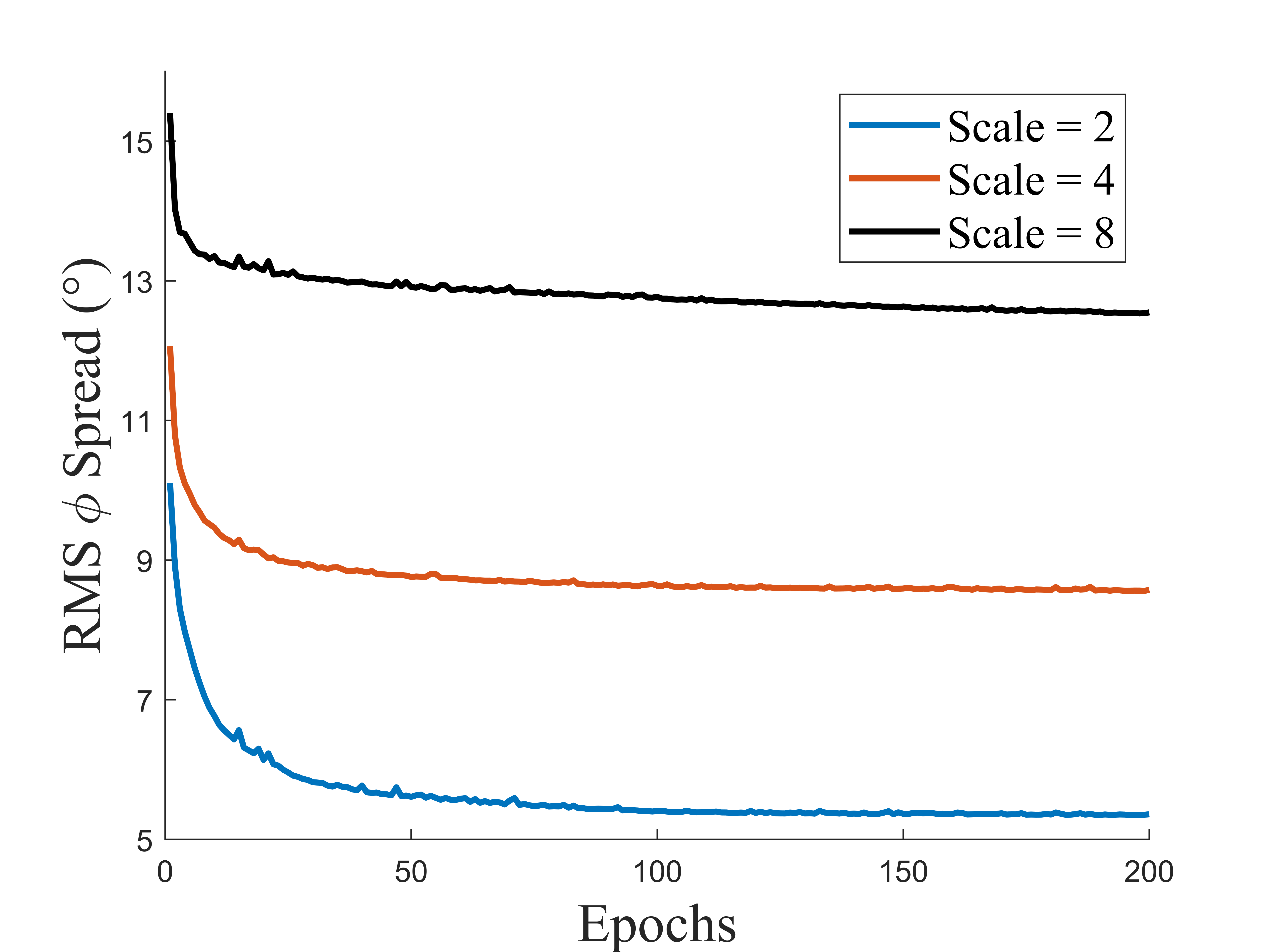}}

	%\qquad
	%让图片换行，
	
	\subfigure[Classification accuracy of LOS/NLOS]{
		\label{LOS_img}
		\includegraphics[width=0.32\linewidth]{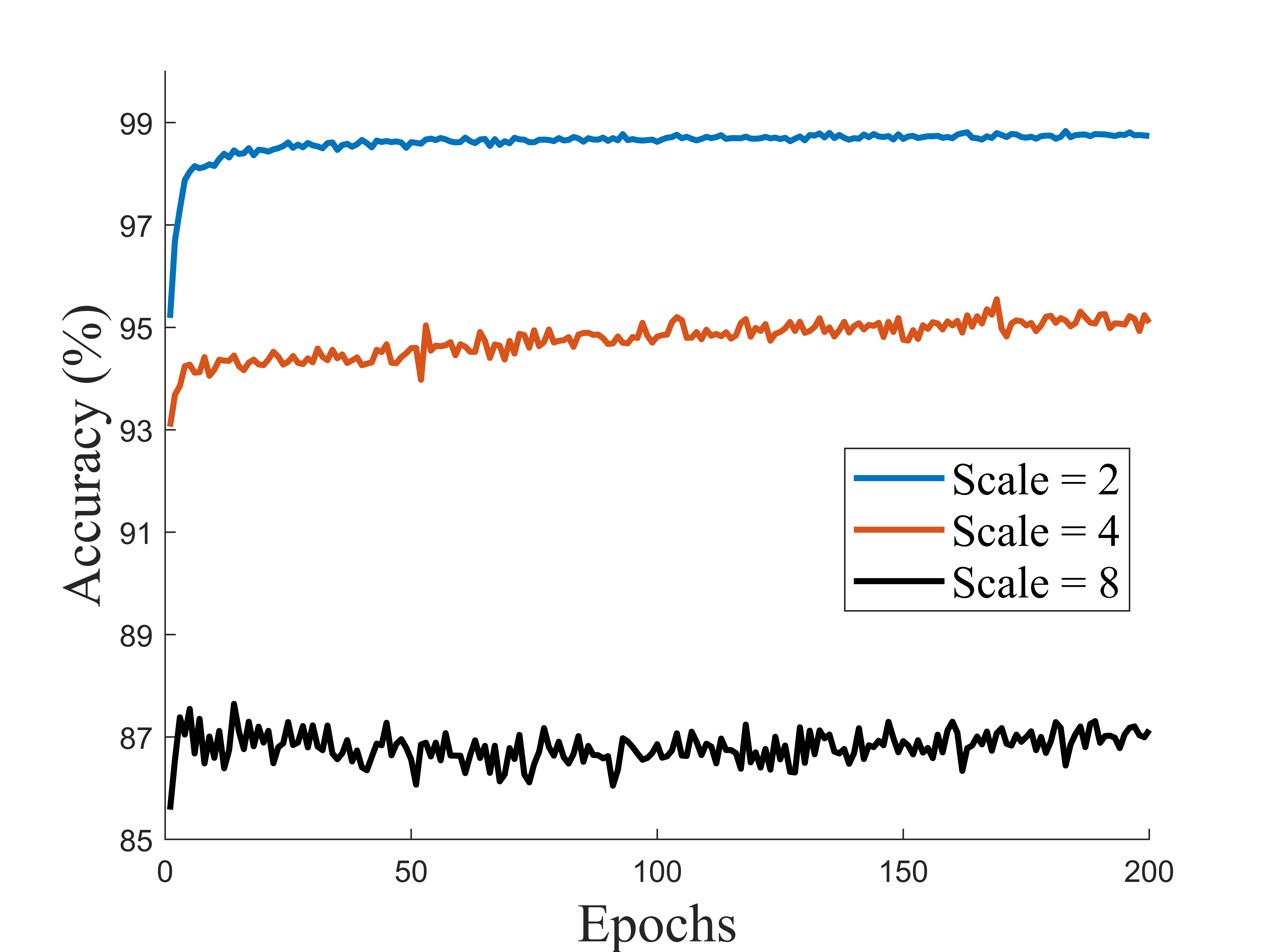}}
	\subfigure[RMS delay spread]{
		\label{DS_img}
		\includegraphics[width=0.32\linewidth]{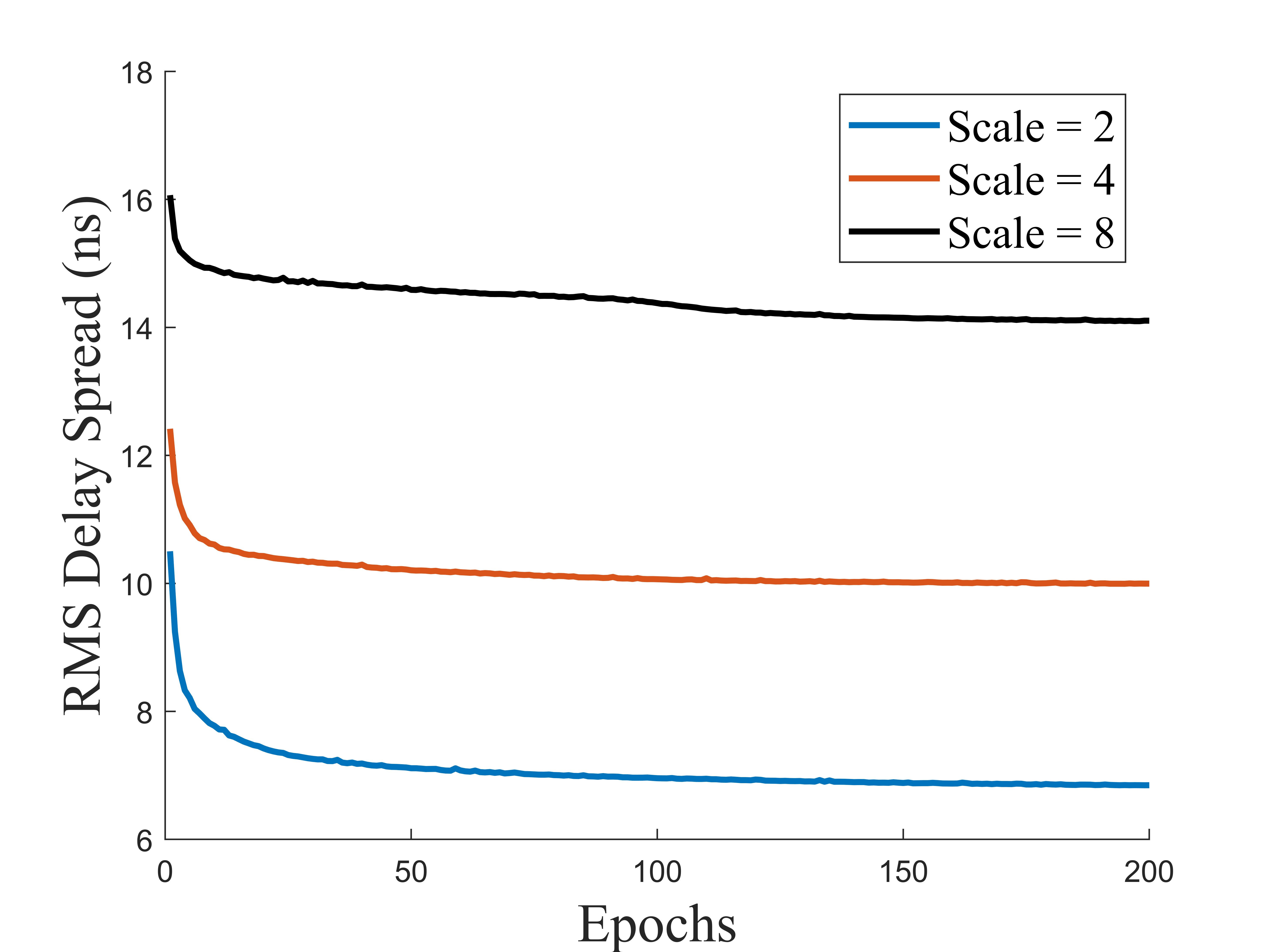}}
	\subfigure[RMS $\theta$ spread]{
		\label{theta_img}
		\includegraphics[width=0.32\linewidth]{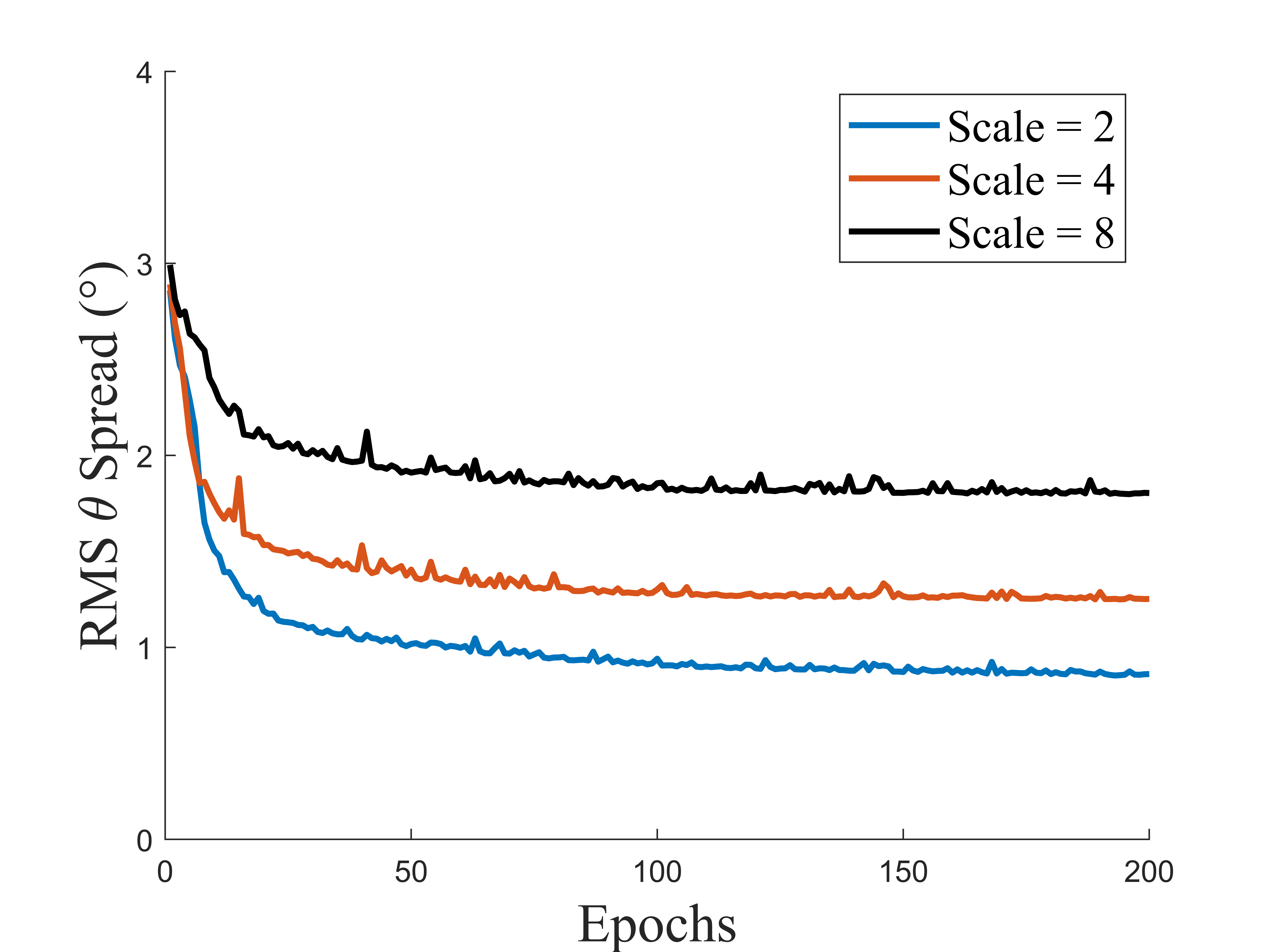}}
	\caption{MAE and classification accuracy of 6 SR targets during training process.}
	\label{trainprocess}
\end{figure*}

\subsection{Backbone Part}

 As mentioned above, \textit{backbone} part of the proposed SR model is regarded as a general feature extractor. The comparison of DL models is summarized in Table \ref{Varioumodel}. Our proposed SR model could achieve 3.26 dB for MAE of PL with scale factor 2. Under the same configuration, several kinds of DL models including ResNet50, vision transformer (ViT)\cite{ViT} and GAN-SR \cite{SRGAN} were tested. Performances of these state-of-the-art DL models are not satisfying. For \textit{backbone} part, ResNet50 is the best in tested DL models but the MAE of PL is around 7-8 dB. After several experiments of ViT by changing the number of transformer encoder and the number of heads in attention layer, the best result of MAE is around 8 dB. Performances of GAN are much worse than CNN and ViT. The best results of GAN is higher than 12 dB for MAE. Moreover, few experiments show that the loss of generator and the output of discriminator (probability that $\hat{I}_{HR}$ is real) converge at the end. Nash equilibrium between generator and discriminator could hardly be reached. For other targets, SR results of these DL models are also much worse than those of the proposed model.
 
The reasons of why state-of-the-art DL models doesn't work well are analyzed. First, the data size of channel characteristics dataset is much smaller than popular computer vision datasets like CIFAR-10, ImageNet and MINST. Second, texture, style and smoothness of images should be considered in image SR but not regarded as objectives in characteristics SR task. As a result, deeper models with larger number of parameters are more liable to over-fitting. The comparison of SR performances and model complexity are presented in Table \ref{Varioumodel}. For the proposed model, both the number of parameters (Params) and floating point of operations (FLOPs) are much smaller than other DL models.
 
 \begin{table}[h]
\centering
\caption{Comparison of deep learning models}
\label{Varioumodel}
\begin{tabular}{l|cccc} 
\toprule
               & \textbf{Proposed Model} & ResNet50 & ViT   & GANSR  \\ 
\midrule
Params / k     & \textbf{4.25}           & 8610     & 13040 & 28670  \\
FLOPs / GMac   & \textbf{0.34}           & 2.58     & 10.33 & 12.42  \\
MAE of PL / dB & \textbf{3.26 (best)}    & 7-8      & 7-8     & $>$12     \\
\bottomrule
\end{tabular}
\end{table}

\subsection{Performance of Proposed Model}

MAE and classification accuracy of 6 SR targets during training process are shown in Fig. \ref{trainprocess}. Clearly for the 6 targets, MAE and classification accuracy of SR are decreasing during training and converge at the end. In \textit{pre-train} stage (1-100 epochs), MAE and classification accuracy decrease very quickly with some minor jitters due to MTL loss. This indicates that the proposed model and MTL loss is effective as a feature extractor to achieve fairly good SR performances with multi inputs and tasks. In \textit{fine-tune} stage (101-200 epochs), few jitters are observed. MAE and classification accuracy decrease slowly to converge at the end so that better SR performances could be achieved in \textit{fine-tune} stage. We also observed that given scale factor as 8, the classification accuracy of LOS/NLOS jitters greatly. We will look into this part in future research. 

In general, the best SR results achieved by the proposed model is demonstrated in Table \ref{MAE error} and Table \ref{STDerror}. Compared with bilinear interpolation, both MAE and STDE of the proposed model is far smaller. Moreover, SR performances of the proposed model deteriorate significantly less than bilinear interpolation when scale factor is relatively larger (4 or 8). 

\subsection{Ablation Study}

Ablation study was conducted to investigate the effectiveness of MTL and techniques in the proposed model. Cumulative SR performance gain of path loss (MAE and STDE) in ablation study are demonstrated in Table \ref{ablationstudy}. \textbf{STL} means the proposed model for single task (PL) during training. \textbf{MTL} represents the proposed model for multi-task learning but without residual connection and \textit{iterative up-and-down} technique. Apparently, SR of single task learning is incomparable to MTL. \textbf{RES} Residual connection and \textit{iterative up-and-down} technique could reduce MAE and STDE for 20\% on average. \textbf{DA} could also enhance SR performance in some extent. It's noticed that the enhancement of  \textbf{DA} and \textbf{RES} decreases much when scale factor are large (4 and 8).

\begin{table}[h]
\centering
\caption{Super resolution performance (MAE) of proposed model and bilinear interpolation} \label{MAE error}
\begin{tabular}{|c|c|c|c|c|c|c|c|} 
\hline
\multicolumn{2}{|c|}{\diagbox{Scale}{Targets}}            & PL                                     & $R_p$                                  & DS                                     & $\phi$                            & $\theta$                           & \begin{tabular}[c]{@{}c@{}}\scriptsize{LOS}/\\ \scriptsize{NLOS}\end{tabular}                              \\ 
\hline
\multirow{2}{*}{2} & Proposed                                  & 3.26                                      & 0.89                                      & 6.84                                      & 5.34                                      & 0.84                                      & 98\%                                     \\ 
\hhline{|~-------|}
                   & {\cellcolor[rgb]{0.89,0.89,0.89}}Bilinear & {\cellcolor[rgb]{0.89,0.89,0.89}}16.58 & {\cellcolor[rgb]{0.89,0.89,0.89}}7.21  & {\cellcolor[rgb]{0.89,0.89,0.89}}13.46 & {\cellcolor[rgb]{0.89,0.89,0.89}}16.60 & {\cellcolor[rgb]{0.89,0.89,0.89}}12.11  & {\cellcolor[rgb]{0.89,0.89,0.89}}85\%  \\ 
\hline
\multirow{2}{*}{4} & Proposed                                  & 4.29                                      & 1.28                                      & 9.99                                      & 8.55                                      & 1.24                                      & 95\%                                      \\ 
\hhline{|~-------|}
                   & {\cellcolor[rgb]{0.89,0.89,0.89}}Bilinear & {\cellcolor[rgb]{0.89,0.89,0.89}}26.76 & {\cellcolor[rgb]{0.89,0.89,0.89}}11.74 & {\cellcolor[rgb]{0.89,0.89,0.89}}20.41 & {\cellcolor[rgb]{0.89,0.89,0.89}}26.53 & {\cellcolor[rgb]{0.89,0.89,0.89}}19.90 & {\cellcolor[rgb]{0.89,0.89,0.89}}72\%  \\ 
\hline
\multirow{2}{*}{8} & Proposed                                  & 5.81                                      & 1.98                                      & 14.09                                      & 12.53                                      & 1.78                                      & 87\%                                      \\ 
\hhline{|~-------|}
                   & {\cellcolor[rgb]{0.89,0.89,0.89}}Bilinear & {\cellcolor[rgb]{0.89,0.89,0.89}}36.77 & {\cellcolor[rgb]{0.89,0.89,0.89}}16.51 & {\cellcolor[rgb]{0.89,0.89,0.89}}27.91  & {\cellcolor[rgb]{0.89,0.89,0.89}}36.92 & {\cellcolor[rgb]{0.89,0.89,0.89}}27.97 & {\cellcolor[rgb]{0.89,0.89,0.89}}65\%  \\ 
\hline

\end{tabular}
\end{table}

\begin{table}[h]
\centering
\caption{Super resolution performance (STDE) of proposed model and bilinear interpolation}
\label{STDerror}
\begin{tabular}{|c|c|c|c|c|c|c|} 
\hline
\multicolumn{2}{|c|}{\diagbox{Scale}{Targets}}            & PL                                     & $R_p$                                   & DS                                     & $\phi$                             & $\theta$                            \\ 
\hline
\multirow{2}{*}{2} & Proposed                                  & 6.09                                     & 2.54                                    & 12.19                                      & 10.23                                      & 1.81                                       \\ 
\hhline{|~------|}
                   & {\cellcolor[rgb]{0.89,0.89,0.89}}Bilinear & {\cellcolor[rgb]{0.89,0.89,0.89}}31.69 & {\cellcolor[rgb]{0.89,0.89,0.89}}13.19 & {\cellcolor[rgb]{0.89,0.89,0.89}}18.18 & {\cellcolor[rgb]{0.89,0.89,0.89}}27.57 & {\cellcolor[rgb]{0.89,0.89,0.89}}24.39  \\ 
\hline
\multirow{2}{*}{4} & Proposed                                  & 7.90                                      & 3.48                                      & 16.82                                      & 16.27                                      & 2.44                                      \\ 
\hhline{|~------|}
                   & {\cellcolor[rgb]{0.89,0.89,0.89}}Bilinear & {\cellcolor[rgb]{0.89,0.89,0.89}}43.24 & {\cellcolor[rgb]{0.89,0.89,0.89}}18.18 & {\cellcolor[rgb]{0.89,0.89,0.89}}24.94 & {\cellcolor[rgb]{0.89,0.89,0.89}}38.41 & {\cellcolor[rgb]{0.89,0.89,0.89}}33.75  \\ 
\hline
\multirow{2}{*}{8} & Proposed                                  & 10.21                                     & 4.85                                      & 20.85                                     & 20.21                                      & 3.20                                      \\ 
\hhline{|~------|}
                   & {\cellcolor[rgb]{0.89,0.89,0.89}}Bilinear & {\cellcolor[rgb]{0.89,0.89,0.89}}51.15 & {\cellcolor[rgb]{0.89,0.89,0.89}}21.82 & {\cellcolor[rgb]{0.89,0.89,0.89}}30.23 & {\cellcolor[rgb]{0.89,0.89,0.89}}46.20 & {\cellcolor[rgb]{0.89,0.89,0.89}}40.42  \\
\hline
\end{tabular}
\end{table}

\begin{table}[h]
\caption{Cumulative super resolution performance gain of path loss  }\label{ablationstudy}
\begin{center}
\begin{threeparttable} 
\begin{tabular}{|l|c|c|c|c|c|c|} 
\hline
\multirow{2}{*}{} & \multicolumn{3}{c|}{\textbf{MAE}} & \multicolumn{3}{c|}{\textbf{STDE}} \\ 
\cline{2-7}
 & scale=2 & scale=4 & scale=8 & scale=2 & scale=4 & scale=8 \\ 
\hline
+DA & +36\% & +26\% & +18\% & +26\% & +14\% & +8\% \\
\hline
+RES & +28\% & +20\% & +13\% & +20\% & +12\% & +7\% \\ 
\hline
MTL & 0 & 0 & 0 & 0 & 0 & 0 \\ 
\hline
STL &-149\%& -170\% & -124\% & -167\% & -221\% & -205\% \\ 
\hline
\end{tabular}
\begin{tablenotes}    %这行要添加， 从这开始
        \footnotesize           %这行要添加
        \item\leftline{+DA: Add data augmentation to +RES.} 
        \leftline{+RES: Add residual connection and \textit{iterative up-and-down} to MTL.}
        \leftline{MTL: The proposed model without RES and DA in training.} 
        \leftline{STL: Single SR task (PL) learning using the proposed model} 
        \leftline{without RES and DA.}
        % \end{footnotesize}
      \end{tablenotes}            %这行要添加

\end{threeparttable}
\end{center}
\end{table}

\section{Conclusion}
In this paper, a novel residual network based MTL model is proposed for SR of wireless channel characteristics. RT simulation was conducted and channel characteristics dataset were constructed based on simulation results. Weighted masks are introduced in loss function which can help better fit randomly distributed building regions. A general MTL model with two stage training methods are proposed. The proposed model could achieve SR results of PL with MAE of 3.26 dB and 98\% classification accuracy of LOS/NLOS areas given scale factor as 2. It also outperforms other state-of-the-art DL models and the reasons are discussed. The proposed model demonstrates huge advantages in channel characteristics SR tasks especially when scale factor is relatively large. Ablation study also proved the necessity of residual connection, \textit{iterative up-and-down} techinque and multi-task learning. In future, we will continue the study of channel characteristics SR problem by refining proposed MTL model on network structure and MTL loss function. Relations between upper bounds of SR performances and scale factor will also be explored.

\section*{Acknowledgment}

This work is supported by the Fundamental Research Funds for the Central Universities 2020JBZD005, National Science Foundation of China under Grant 61901029, Beijing Natural Science Foundation L212029, the State Key Laboratory of Rail Traffic Control and Safety (Contract No. RCS2020ZZ005), Beijing Jiaotong University, Ministry of Education of China under Grant (8091B032123), ZTE Corporation and State Key Laboratory of Mobile Network and Mobile Multimedia Technology. 

\bibliographystyle{IEEEtran}
\small\bibliography{ref}

\end{document}